\documentclass[twocolumn,showpacs,preprintnumbers,amsmath,amssymb]{revtex4-1}

\usepackage{epsfig}
\usepackage{dcolumn}
\usepackage{bm}
\usepackage{xcolor}
\usepackage{amsmath}
\usepackage{ulem}
\usepackage{float}
\usepackage{siunitx}

\begin{document}

\preprint{\today} 

\title{Measurement of the ratio of the scalar polarizability to the vector polarizability for the $6s ^2S_{1/2} \rightarrow 7s ^2S_{1/2}$ transition in cesium}

\author{Jonah A. Quirk$^{1,2}$,  Carol E. Tanner$^3$ and D. S. Elliott$^{1,2,4}$}

\affiliation{%
   $^1$Department of Physics and Astronomy, Purdue University, West Lafayette, Indiana 47907, USA\\
   $^2$Purdue Quantum Science and Engineering Institute, Purdue University, West Lafayette, Indiana 47907, USA\\
   $^3$Department of Physics and Astronomy, University of Notre Dame, Notre Dame, Indiana 46556, USA\\
    $^4$School of Electrical and Computer Engineering, Purdue University, West Lafayette, Indiana 47907, USA
   }

\date{\today}

\begin{abstract}
We report measurements of the ratio of the scalar polarizability $\alpha$ to the vector polarizability $\beta$ for the $6s ^2S_{1/2} \rightarrow 7s ^2S_{1/2}$ transition in atomic cesium.  
These measurements are motivated by a discrepancy between the values of the vector transition polarizability as determined using two separate methods. In the present measurement, we use a two-pathway, coherent-control technique in which we observe the interference between a two-photon interaction driven by infrared light at 1079 nm and a linear Stark-induced interaction driven by the mutually-coherent second harmonic of this infrared beam at 540 nm.  The result of our measurements is $\alpha/\beta = -9.902 \: (9)$, in good  agreement with the previous determination of this ratio.  This measurement, critical to the study of atomic parity violation in cesium, does not reduce the discrepancy between the two methods for the determination of the vector polarizability $\beta$ for this transition. 
\end{abstract}

\maketitle 

\section{Introduction}\label{sec:introduction}
In atomic parity violation (APV) investigations, researchers carry out precise measurements of the strength of optical interactions that are allowed only because of the weak force interaction between the nucleons and the electrons, as mediated by the exchange of the neutral $Z^0$ vector boson.  Measurements of the strength of these extremely weak transitions, combined with precise atomic structure calculations, allow for a precise determination of the weak charge $Q_w$, as well as $\sin^2 \theta_w$, where $\theta_w$ is the Weinberg mixing angle.  Any discrepancy between the measured value of $Q_w$ or $\sin^2 \theta_w$ and the standard model values can be an indication of physics beyond the standard model.  It is therefore of great interest to push the accuracy of these measurements to new limits. 

Due to the minute oscillator strength of these weak-force-induced optical transitions, direct APV measurements are not feasible. To enable detection, all APV measurements to date are based upon an interference between the amplitude $A_{\rm PNC}$ of the weak transition and the amplitude of a stronger optical interaction, such as a Stark-induced interaction ($A_{\rm St}$) or a magnetic dipole interaction ($A_{\rm M1}$). 
For example, in their APV measurement on the $6s ^2S_{1/2} \rightarrow 7s ^2S_{1/2}$ transition in cesium, which is to date the most precise APV measurement in any element, Wood \textit{et al.}~\cite{WoodBCMRTW97} reported their measurement results in terms of $\mathcal{E}_{\rm PNC}/\beta$, where $\mathcal{E}_{\rm PNC}$ is the weak-force-induced electric dipole moment for the transition, and $\beta$ is the vector transition polarizability. A proper evaluation of $\mathcal{E}_{\rm PNC}$ therefore requires a precise value for the vector polarizability, $\beta$.  

Until 2019, the most precise value of $\beta$ was based upon a theoretical value for the hyperfine-changing component of the magnetic dipole matrix element $M1^{\rm hf}$~\cite{DzubaF00}, and a laboratory determination of the ratio $M1^{\rm hf} / \beta$ \cite{BennettW99}. The value of $\beta$ determined in this way is $\beta_{\rm M1} = 26.957 \: (51) \: a_0^3$, with a precision of 0.19\%.  The subscript `M1' (or later `$\alpha$') on $\beta$ indicates the method of determination.

The alternative method to determine $\beta$ uses a sum-over-states calculation of the scalar transition polarizability $\alpha$~\cite{BlundellSJ92,SafronovaJD99, VasilyevSSB02, DzubaFG02, TanXD2023}, combined with a measurement of the ratio $\alpha / \beta$ ~\cite{ChoWBRW1997}. This approach requires precise measurements of or theoretical values for the reduced electric dipole (E1) matrix elements $\langle np_J || r || ms\rangle$ with $m = 6$ or $7$, $n \ge 6$ and $J = 1/2$ or $3/2$.  (We will use this abbreviated state notation $ms$ for $ms ^2S_{1/2}$ and $np_J$ for $np ^2P_J$.)  Many of these matrix elements have been measured to great precision over the past thirty years~\cite{BouchiatGP84, TannerLRSKBYK92, YoungHSPTWL94, RafacT98, RafacTLB99, BennettRW99, VasilyevSSB02, DereviankoP02a, AminiG03, BouloufaCD07, SellPEBSK11, ZhangMWWXJ13, antypas7p2013, Borvak14, PattersonSEGBSK15, GregoireHHTC15}, and in the last six years, our group has undertaken and completed high-precision measurements for six of the eight most significant E1 matrix elements~\cite{TohJGQSCWE18, TohDGQSCSE19, DamitzTPTE18a, quirk2024starkshift}. Prior to 2023, the discrepancy between the two methods for determining $\beta$ was $0.67\%$, greater than the sum of their uncertainties. Recent exhaustive theoretical calculations of reduced E1 matrix elements $\langle np_J || r || ms\rangle$ by Tan \textit{et al.}~\cite{TanD2023} along with our recent measurement of the Stark shift of the $7s$ state \cite{quirk2024starkshift} reduced this discrepancy to $0.29\%$, slightly larger than either uncertainty. This determination utilized a mix of high precision measurements of reduced E1 matrix elements along with some theoretical calculations where experimental results are missing or imprecise to determine $\alpha$. This result combined with the measurement of $\alpha/\beta$ \cite{ChoWBRW1997} yields $\beta_{\alpha}=27.043\:(36)\:a^3_0$ \cite{quirk2024starkshift}. Tan \textit{et al.}~also report a determination of $\beta_{\alpha}= 26.887\:(38)\:a_0^3$ \cite{TanXD2023} utilizing only theoretical calculations of reduced E1 matrix elements for the determination of $\alpha$ and the measured ratio $\alpha/ \beta$ \cite{ChoWBRW1997}. 
%

To investigate a possible cause for this $0.29 \%$ discrepancy between values of $\beta$, we have carried out a new precision measurement of the ratio $\alpha/\beta$ for the cesium $6s \rightarrow 7s$ transition, which we report in this work.  Our technique is based upon two-color, two-pathway coherent control, in which we excite the transition via two coherent optical interactions; two-photon absorption of a laser field at a wavelength of 1079 nm and Stark-induced single-photon absorption of the second-harmonic light at a wavelength of 540 nm.  We describe the coherent control process in Sec. II below.  In Sec.~\ref{sec:Measurement}, we outline the experimental technique, followed by a discussion of the pumping efficiency measurement of the atomic beam.  In Sec.~\ref{sec:SysContr}, we describe possible systematic effects and methods for reducing them below $0.1\%$ of $\beta$.  We discuss our results in Sec.~\ref{sec:Results} and conclusions in Sec.~\ref{sec:Conclusion}.

\begin{figure*}
    \includegraphics[width=0.9\textwidth]{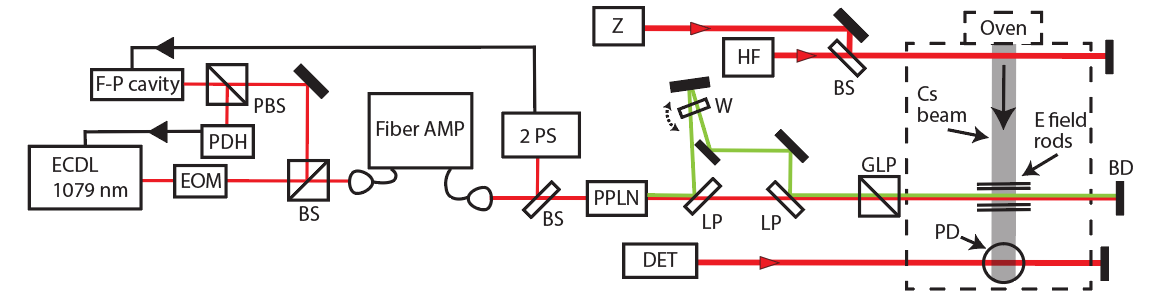} 
    \caption{An experimental diagram of the $\alpha/\beta$ measurement. The 1079 nm ECDL is stabilized to a Fabry-Perot cavity (F-P cavity) and amplified in a fiber amplifier. This amplified 1079 nm beam is frequency doubled in a periodically-poled lithium niobate crystal (PPLN). The relative phase of the second harmonic at 540 nm and the fundamental at 1079 nm is varied in a Mach-Zehnder interferometer where one arm contains a galvanometer mounted window (W). The 540 nm and 1079 nm beams are recombined and polarized with a Glan-laser polarizer (GLP) before being focused and directed into the vacuum chamber. The following elements are labeled as; 2 PS - two-photon stabilization, ECDL - external cavity diode laser, PDH - Pound Drever Hall locking system, BD -beam dump, BS - beam splitter, PBS - polarizing beam splitter, DET - detection laser, Z - Zeeman laser, HF - hyperfine laser, LP - long pass filter, PD - photodetector, and EOM - Electro-optic modulator. }
    \label{fig:exptdiagram}
\end{figure*}

\section{Two-Pathway Coherent Control}\label{sec:TPCC}
In two-pathway coherent control, one exploits the interference between optical interactions that are driven by two distinct, but mutually-coherent, laser fields.  We have used this technique previously to measure $M1/\beta$ of the $6s \rightarrow 7s$ transition in cesium~\cite{antypasm12013,antypase2014}, where $M1$ is the transition magnetic dipole moment.  
In the present work, we interfere a strong two-photon allowed E1 excitation, driven by an infrared (IR) laser at 1079 nm, with a weaker Stark induced one-photon excitation  driven by the second-harmonic radiation at 540 nm. The net excitation rate $\mathcal{W}$ of the $7s$ state is governed by the square of the sum of the transition amplitudes, $A_{\rm 2p}$ for the former and $A_{\rm St}$ for the latter.  Then by Fermi's Golden Rule, the transition rate is
\begin{displaymath}
  \mathcal{W} = \frac{2 \pi}{\hbar} \left|\rule{0in}{0.15in} A_{\rm 2p} + A_{\rm St} \right|^2 \tilde{\rho}_{7s}(E),
\end{displaymath}
where $\tilde{\rho}_{7s}(E)$ is the density of states of the $7s$ state.  
When the laser is tuned to line center and the linewidth is lifetime limited, the transition rate simplifies to
\begin{displaymath}
  \mathcal{W} = \frac{4}{\hbar^2 \Gamma} \left|\rule{0in}{0.15in} A_{\rm 2p} + A_{\rm St} \right|^2 ,
\end{displaymath}
where $\Gamma$ is the decay rate of the $7s$ state.   
Expanding the sum, the transition rate $\mathcal{W}$ becomes
\begin{equation}\label{eq:rate}
  \mathcal{W} = \frac{4}{\hbar^2 \Gamma} \left\{  \rule{0in}{0.15in} \left| A_{\rm 2p} \right|^2 + \left( A_{\rm 2p} A_{\rm St}^{\ast} + A_{\rm 2p}^{\ast} A_{\rm St} \right) + \left| A_{\rm St} \right|^2 \right\}.
\end{equation}
The excitation rate consists of d.c.\ terms due to $A_{\rm 2p}$ and $A_{\rm St}$ by themselves, plus a beat term resulting from the interference between the two amplitudes $A_{\rm 2p}$ and $A_{\rm St}$.  The beat signal can be modulated by varying the phase difference between the amplitudes $A_{\rm 2p}$ and $A_{\rm St}$.  

The two-photon amplitude for the $6s \rightarrow 7s$ transition driven by a single-frequency field scales as the square of the field amplitude $\boldsymbol{\varepsilon}^{\omega}$ of the 1079 nm laser field,
\begin{equation}
 A_{\rm 2p} = \tilde{\alpha} \boldsymbol{\varepsilon}^{\omega} \cdot \boldsymbol{\varepsilon}^{\omega} \delta_{F,F^{\prime}}  \delta_{m,m^{\prime}} e^{-2i \phi^{\omega}},
\end{equation}
where we define $\phi^{\omega}$ to be the phase of the 1079~nm beam and $\tilde{\alpha}$ is the scalar polarizability for the 2-photon excitation. The phase factor must be included here, as it becomes relevant in the interference effect.  The $\delta$ functions represent the selection rules that only $\Delta F=0$, $\Delta m=0$ transitions are allowed~\cite{bonin1984two}. $F$ and $m$ ($F^{\prime}$ and $m^{\prime}$) are the quantum numbers indicating the total angular momentum, electronic plus nuclear spin, of the ground $6s$ state (excited $7s$ state) and its projection on the $z$ axis, respectively.

The Stark amplitude $A_{\rm St}$ is linear in the applied static electric field $\mathbf{E}$ and the field amplitude of the 540 nm beam $\boldsymbol{\varepsilon}^{2 \omega}$, and depends upon the relative orientation between $\mathbf{E}$ and $\boldsymbol{\varepsilon}^{2 \omega}$. The amplitude for the Stark-induced transition can be written~\cite{GilbertW1986}
\begin{eqnarray}
\label{eq:stark}
  A_{\rm St} &=& \left\{ [\alpha \mathbf{E} \cdot \boldsymbol{\varepsilon}^{2 \omega} \delta_{F,F^{\prime}} + i \beta  \left( \mathbf{E} \times \boldsymbol{\varepsilon}^{2 \omega} \right)_z C_{F,m}^{F^{\prime},m^{\prime}}] \delta_{m,m^{\prime}} \right. \\
 && \hspace{-0.4in} \left. +[\pm i \beta  \left( \mathbf{E} \times \boldsymbol{\varepsilon}^{2 \omega} \right)_x - \beta \left( \mathbf{E} \times \boldsymbol{\varepsilon}^{2 \omega} \right)_y] C_{F,m}^{F^{\prime},m^{\prime}} \delta_{m,m^{\prime} \pm 1} \right\} e^{-i \phi^{2\omega}}. \nonumber
\end{eqnarray}
$\alpha$ ($\beta$) is the scalar (vector) transition polarizability, which parameterizes the transition amplitude when $\mathbf{E}$ and $\boldsymbol{\varepsilon}^{2 \omega}$ are parallel (perpendicular) to one another. We define $\phi^{2\omega}$ to be the phase of the 540~nm beam, and the  coefficients $C_{F,m}^{F^{\prime},m^{\prime}}$ are derived from Clebsch-Gordon coefficients and tabulated for this transition in Ref.~\cite{GilbertW1986}. The relevant coefficients for the present study are $C_{3,m}^{3,m} = -m/4$, which for $m=\pm3$ is $\mp3/4$.
 
In our measurements, the two-color (1079 nm and 540 nm) laser field intersects an atomic cesium beam perpendicularly inside a vacuum chamber. Both frequency components of the optical field are linearly polarized.  We apply a static magnetic field $\mathbf{B} \approx$ 8.8 G that is closely aligned with $\hat{\mathbf{k}}$, the direction of propagation of both laser fields, which we use to define the $z$-axis of our coordinate system. With the laser tuned to a $\Delta F = 0$, $\Delta m = 0$ transition, the Stark amplitude simplifies to 
\begin{equation}
 A_{\rm St} = E \varepsilon^{2 \omega}  \left[ \alpha  \cos \theta + i \beta C_{F,m}^{F',m'} \sin \theta  \right]  e^{-i\phi^{2 \omega}}   ,
\end{equation}
where $\theta$ is the angle between the static field $\mathbf{E}$ and the polarization direction of the green beam $\boldsymbol{\varepsilon}^{2 \omega}$. We scan the relative phase of the 2-photon and Stark laser fields, $\Delta \phi=\phi^{2\omega}-2\phi^{\omega}$, and measure the modulation amplitude when the polarization is parallel and perpendicular to the static electric field. The ratio of these measurements yields,
\begin{equation}\label{eq:ReqslphaoverbetaC}
    \mathcal{R}=\frac{\alpha}{\beta C_{F,m}^{F',m'}}
\end{equation}
when the polarization is perfectly parallel or perpendicular to the electric field and the population is fully prepared in a single angular momentum substate $m$.  When the atomic preparation is not complete, the ratio of amplitudes can be shown to be 
\begin{equation}\label{eq:Reqslphaoverbetam}
    \mathcal{R}= \frac{4}{\langle m \rangle } \frac{\alpha}{\beta},
\end{equation}
where $ \langle m \rangle$ is the average value of $m$ for ground state atoms in the interaction region.

We write the imperfect polarization of the $540$ nm beam as $\boldsymbol{\varepsilon}^{2\omega}= \hat{\mathbf{y}} \varepsilon_{y}+ \hat{\mathbf{x}} (\varepsilon'_{x}+i\varepsilon_{x}'')$, where $\varepsilon_{y}$ represents the primary component of the green ($\varepsilon^{2\omega}$) beam, $\varepsilon'_{x}$ a slight rotation of the polarization from the intended direction, and $\varepsilon''_{x}$ the slight amount of circular polarization remaining in the $\varepsilon^{2\omega}$ beam. The measured ratio $\mathcal{R}$ is modified to,
\begin{equation}
\begin{split}
\label{eq:ratio_corrected}
\mathcal{R}_{\pm}=\frac{4}{| \langle m \rangle | } \frac{\alpha}{\beta} &\Bigg(1\mp\frac{4}{ | \langle m \rangle |} \frac{\alpha}{\beta} \frac{\varepsilon_x''}{\varepsilon_y}\\&+
\left(\frac{4}{ | \langle m \rangle |} \frac{\alpha}{\beta}\right)^2 \left[\left(\frac{\varepsilon_x''}{\varepsilon_y}\right)^2-\frac{1}{2}\left(\frac{\varepsilon'_x}{\varepsilon_y}\right)^2\right]  \Bigg).
\end{split}
\end{equation}
%
Here the $\pm$ in $\mathcal{R}_\pm$ refers to whether the atoms are initially pumped into the $m = +3$ or $-3$ Zeeman sublevel. Under an ``$m$'' reversal, the second term in Eq.~(\ref{eq:ratio_corrected}) changes sign. 
Although the higher order terms do not change sign and $\alpha/\beta $ is $\sim9.9$, we can we adjust $\varepsilon'_{x}/\varepsilon_{y}$ and $\varepsilon''_{x}/\varepsilon_{y}$ to be less than \num{1e-3} and \num{5e-4} respectively, making these corrections negligibly small. To extract $\alpha / \beta$, we carry out successive measurements of $\mathcal{R}_{+}$ and $\mathcal{R}_{-}$, and average these results. This is the basis of the measurement reported in this work.


\section{Experimental Configuration and Measurement Procedure}\label{sec:Measurement}


\begin{figure*}
 	  \includegraphics[width=0.235\textwidth]{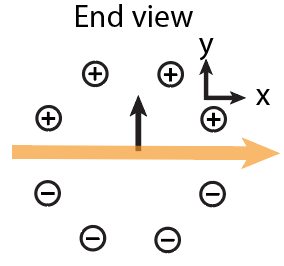} \includegraphics[width=0.235\textwidth]{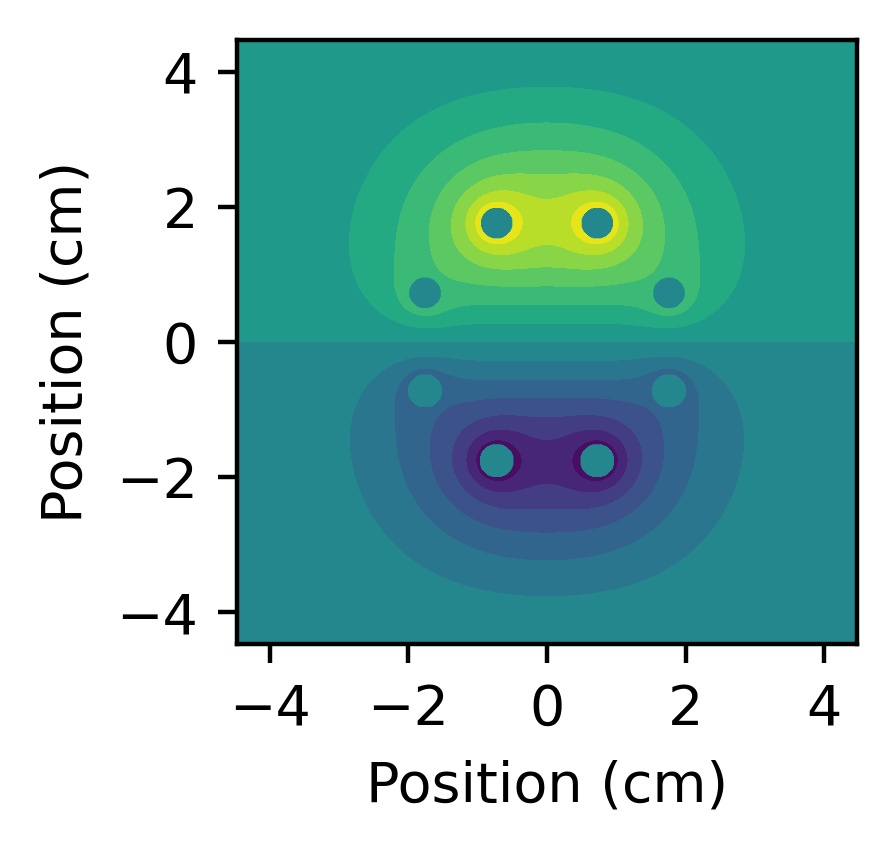}  \includegraphics[width=0.235\textwidth]{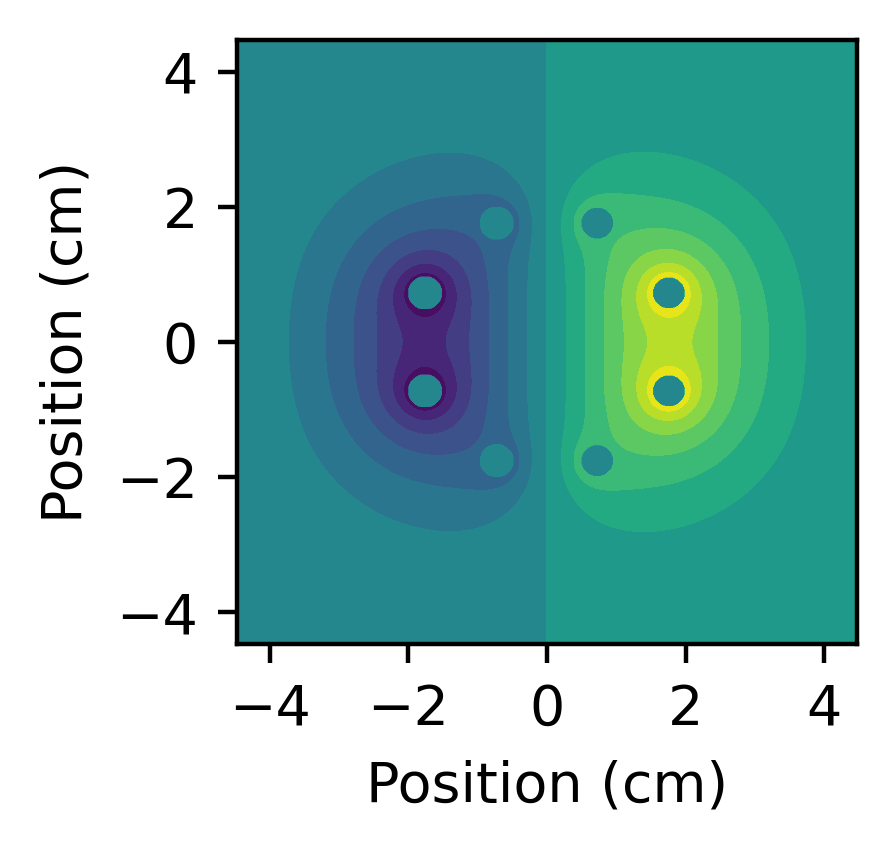} \includegraphics[width=0.235\textwidth]{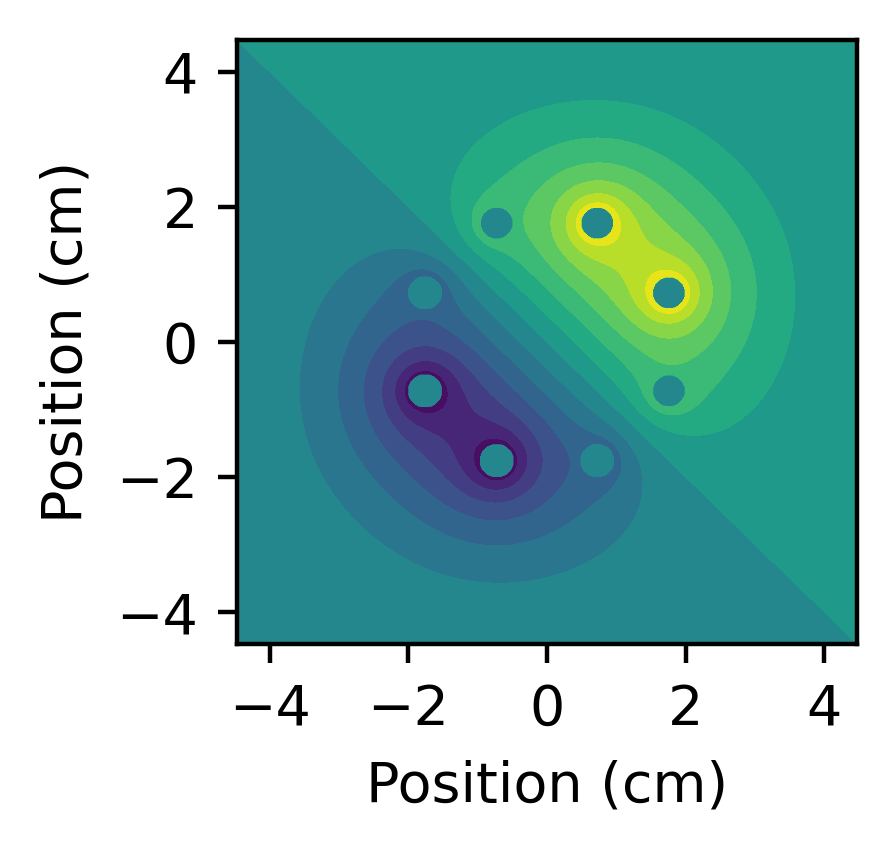} \\ \vspace{-0.59in} \hspace{-6.6 in}{\color{black}(\textbf{a}) } \hspace{0.0in} \\
    \vspace{-1\baselineskip} \hspace{-1.65 in}{\color{white}(\textbf{b}) $\angle \mathbf{E} = 0^{\circ}$ } \hspace{0.0in}\\
    \vspace{-1\baselineskip} \hspace{1.92 in}{\color{white}(\textbf{c}) $\angle \mathbf{E} = +90^{\circ}$ } \hspace{0.0in}\\
    \vspace{-1\baselineskip} \hspace{5.37 in}{\color{white}(\textbf{d}) $\angle \mathbf{E} = +45^{\circ}$ } \hspace{0.0in} \\\vspace{0.35in}
  \caption{(a) Diagram illustrating the interaction region geometry and (b-d) modeled electric potentials generated by the circular arrangement of eight biased conducting rods. In (a) the large, orange arrow depicts the cesium beam and the small, black arrow depicts the laser polarization. The laser propagates into the page in each illustration (a-d).  The rod assembly allows for rapid, reproducible rotation of the static electric field ($\mathbf{E}$) while the polarization ($\boldsymbol{\varepsilon}^{2 \omega}$) remains fixed. 
  The angle labeled in the figures ($\angle \mathbf{E}$) is specified relative to vertical.}
        \label{fig:Electricfieldrod}
\end{figure*}

An illustration of the experimental configuration is depicted in Fig.~\ref{fig:exptdiagram}. The primary laser source for these measurements is a commercial external-cavity-diode laser (ECDL) tuned to the frequency of the two-photon $6s, F=3 \rightarrow 7s, F=3$ transition in cesium.  This laser generates 70 mW of infrared light at a wavelength of 1079 nm and its frequency is locked to a resonance of an invar-mounted, 15-cm long, Fabry-Perot cavity to minimize short-term drifts.  The cavity length is locked to the Doppler-free two-photon absorption resonance signal produced with a cesium vapor cell and photomultiplier.  We amplify the output of the 1079 nm laser in a fiber amplifier to a power of 10 W with a power stability of $0.5\:\%/$8 hr, and frequency double this beam in a periodically-poled lithium niobate (PPLN) frequency doubling crystal.  The output power at 540 nm is 1150 mW, with a peak-to-peak variation of $<0.4\:\%/\rm{hr}$.  We separate the 1079 nm and 540 nm beams with a dichroic beamsplitter, phase delay the green beam in a galvanometer-mounted optical window, and recombine the beams on a second dielectric beamsplitter.  After recombination, we carefully overlap the two beams, and weakly focus them onto the atomic cesium beam inside a vacuum chamber, crossing at nearly a right angle. The waist diameter of the 540 nm beam as it intersects the atomic beam is $\approx$700 $\mu$m and the waist diameter of the 1079 nm beam is $\approx$920 $\mu$m. These waist diameters equate to maximum beam intensities of 150 W/cm$^2$ for the 540 nm beam and 225 W/cm$^2$ for the 1079 nm beam.

The atom beam is generated by an effusive oven, and collimated using a packed array of 0.8 mm inner diameter, 1 cm long stainless steel capillary tubes. These tubes are packed into the nozzle opening, 8 mm high and 12 mm wide.  We optically pump the atoms into a single hyperfine component of the ground state, $F=3$ and $m=\pm 3$, by tuning two preparation laser beams to various hyperfine components of the $6s \rightarrow 6p_{3/2}$ transition. This technique is well described in \cite{Woodthesis1996, dionysiosthesis2013}.

The interaction region for this measurement is defined by the intersection of the atomic beam and the two-color laser field.  We apply a static electric field to the atoms in the interaction region using an assembly of eight parallel copper rod electrodes that is co-axial with the 1079 nm and 540 nm beams, as illustrated in Fig.~\ref{fig:Electricfieldrod}. 
The rods are arranged in a ring configuration, with each rod parallel to the laser propagation direction $\hat{\mathbf{k}}$. Each rod has a diameter of 4.8 mm, and the radius of the ring pattern is 18 mm. Careful choice of the bias voltages applied to each rod allows us to create a uniform electric field in the center of the configuration, which coincides with the interaction region.  To rotate the direction of $\mathbf{E}$, we have constructed a switching circuit consisting of solid state relays which rotates the bias voltages applied to each of the electrodes.  We show color plots of the electric potential for three configurations of potentials applied to the rods in Fig.~\ref{fig:Electricfieldrod}.  We have modeled the electric potential in the region bounded by the electrodes, and find that the electric field, with magnitude 429.7 V/cm, is uniform in direction and magnitude to within 20 mV/cm over the 2 mm diameter volume surrounding the interaction region.  We chose this method to vary the angle $\theta$ between the laser field $\boldsymbol{\varepsilon}^{2 \omega}$ and the static field $\mathbf{E}$ in order to avoid slight changes in the spatial overlap of the two laser beams and/or variations in the optical power or polarization quality of the green beam that might accompany rotation of the laser polarization $\boldsymbol{\varepsilon}^{2 \omega}$. 

\begin{figure}
 \includegraphics[width=0.45\textwidth]{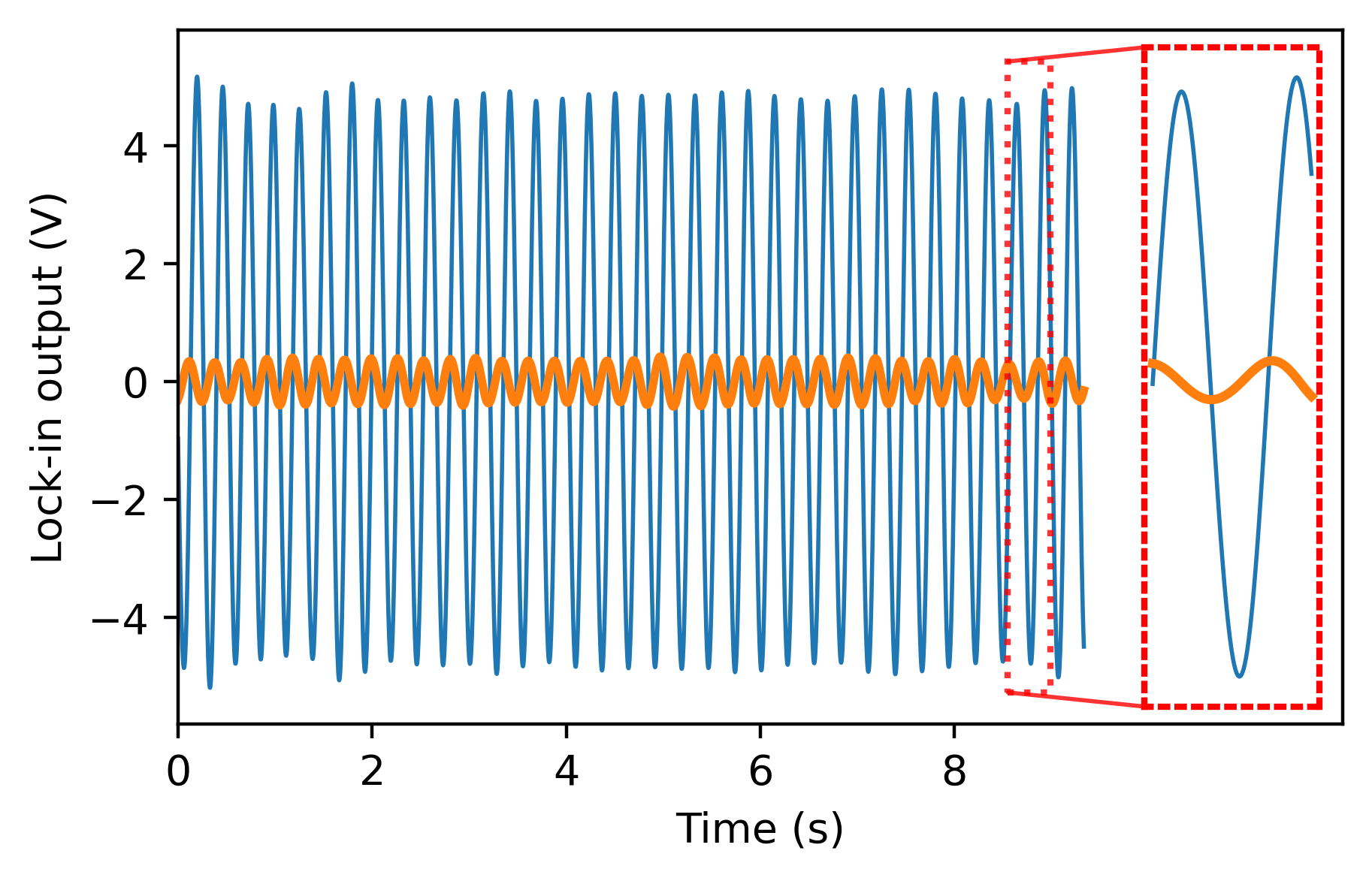}
 \vspace{-1\baselineskip}
 \caption{Representative examples of the bandpass-filtered output of the lock-in amplifier, showing the sinusoidal modulation vs.~time, as the optical phase difference $\Delta \phi$ is scanned. Here atoms are prepared into the $m = +3$ Zeeman sublevel, so $C_{F,m}^{F'm'}=-3/4$.  The larger (thin blue) trace demonstrates the $\alpha$ interference with an electric field parallel to the static polarization and smaller (thick orange trace) illustrates $\beta$ interference. The inset plot (dotted red section) is horizontally stretched to highlight the phase difference between the $\alpha$ and $\beta$ interference. This shift is consistent with a negative value for the ratio of $\alpha/\beta$.}
 	  \label{fig:lockinoutput}
\end{figure}

Atoms that participate in the $6s, F=3 \rightarrow 7s, F=3$ transition may decay down to the $6s, F=4$ level.  When these atoms reach the detection region, 20 cm downstream from the interaction region, they are excited through a cycling transition via a detection laser tuned to the $6s, F=4 \rightarrow 6p_{3/2}, F=5$ transition. This process scatters many photons which we collect on a large area photodetector. The photocurrent from this photodetector is amplified with a transimpedance amplifier (TIA) of gain 20 M$\Omega$. To vary the phase difference $\Delta \phi$, we linearly ramp (12 s period) while slightly modulating (150 Hz) the phase of the single-photon beam using the galvanometer mounted window. This linear ramp causes the relative phase, $\Delta \phi$, to scan at a rate of $\Omega=3.8$ Hz. We use the 150 Hz modulation as the reference for the lock-in-amplifier to mix down this low frequency modulation, $\Omega$. The TIA output is sent to the input of the lock-in amplifier and the output of the lock-in amplifier is digitally bandpass-filtered and recorded. The bandpass filter is centered on the phase scanning frequency, $\Omega$, and reduces low frequency drifts and higher frequency noise that is near the modulation frequency, 150 Hz.
We scan through $>36$  cycles, reset the galvanometer,  rotate the orientation of the static electric field, and then scan through $>36$ more cycles.  We show sample phase scans of the output of the lock-in-amplifier with $\mathbf{E} \parallel \boldsymbol{\varepsilon}^{2 \omega} $ ($\alpha$, blue) and $\mathbf{E} \perp \boldsymbol{\varepsilon}^{2 \omega} $ ($\beta$, orange) in Fig. \ref{fig:lockinoutput}.

The first 2.6 s of the scan are removed to allow the band-pass filter time  to stabilize. We cut the resultant 36 cycle scan into 12 sections to avoid the effects of small phase variations present during long scans. Each section is fit to a sinusoid to determine its amplitude. The average amplitude of each fit in a single scan is recorded as well as the standard error of the fitted amplitudes.  The $\alpha$ and $\beta$ (i.e.\ the blue and orange traces, respectively, in Fig. \ref{fig:lockinoutput}) interference amplitudes and their respective uncertainties  are combined to obtain a ratio and an uncertainty. This process is repeated while reversing the Zeeman pumping to change the sign of the $C_{F,m}^{F',m'}$ coefficient and cancel out systematic errors due to the small circular polarization contribution to the $\beta$ signal.  We make $\approx$ 160 ratio measurements and combine these ratios weighted by the inverse of their uncertainties squared ($1/\sigma^2$) to attain a final ratio and uncertainty.

 \begin{figure}
 \begin{centering}
 	  \includegraphics[width=0.45\textwidth]{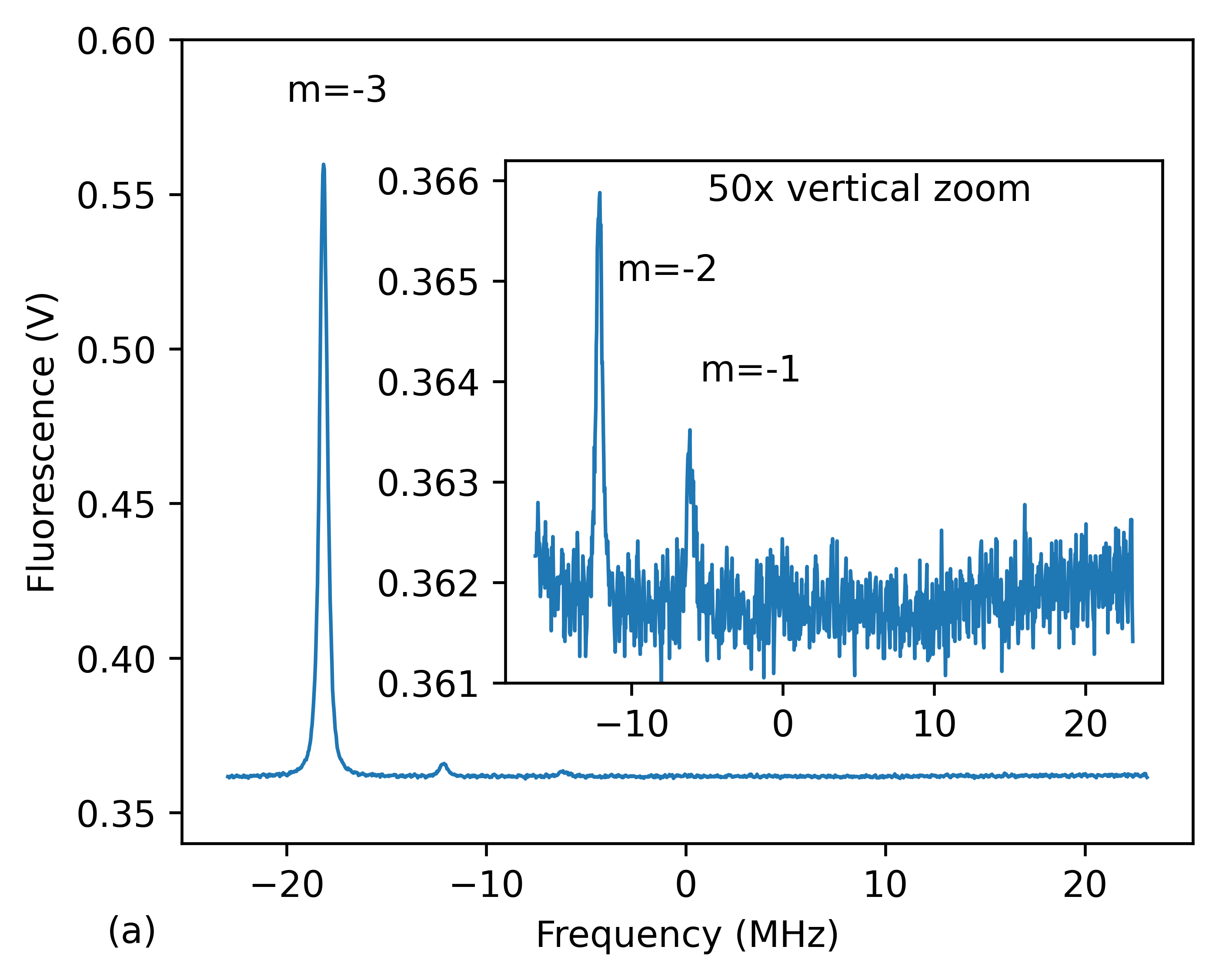} \\\includegraphics[width=0.45\textwidth]{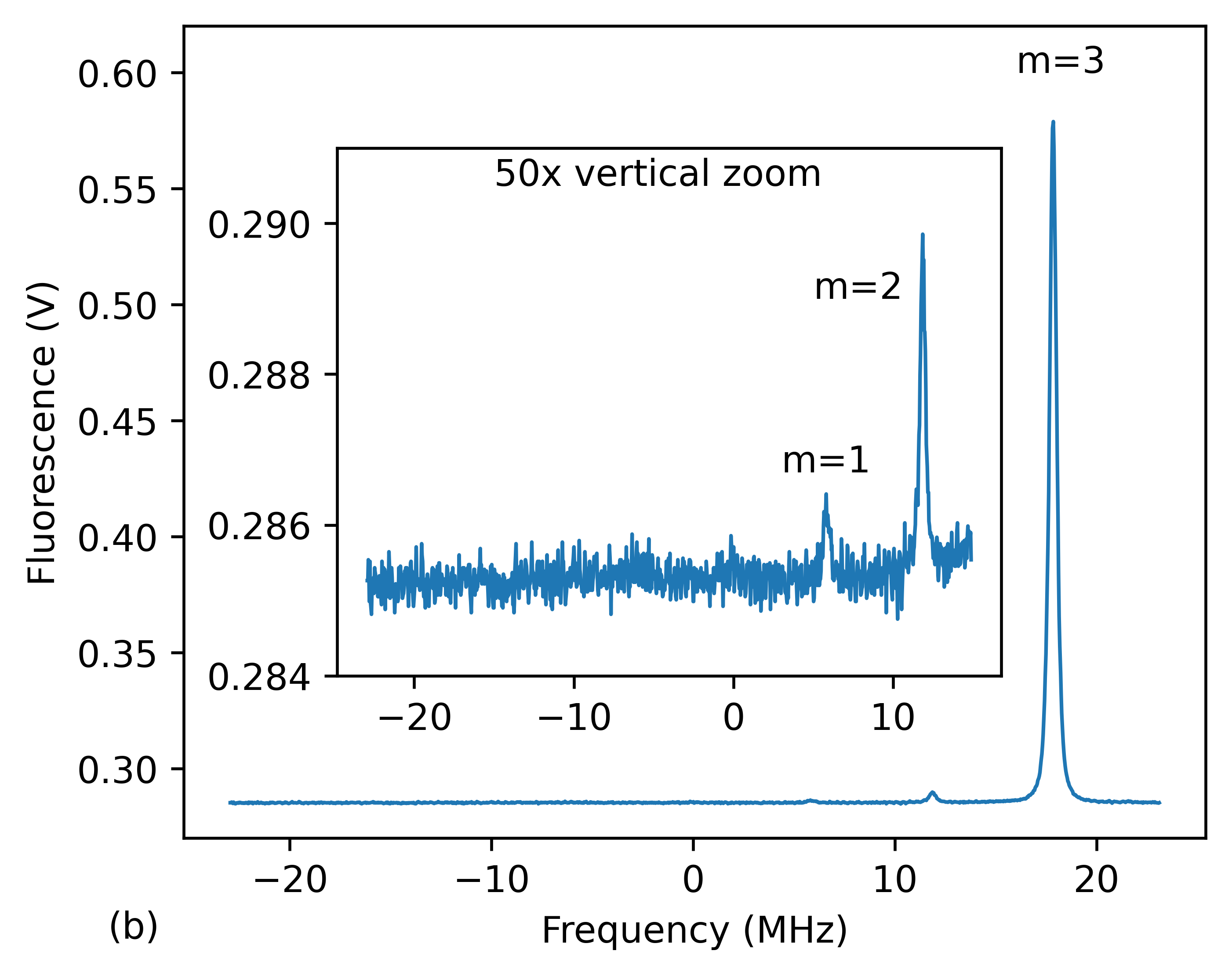}
  \caption{Raman spectra illustrating the effect of optical pumping of the cesium atoms to the most extreme magnetic Zeeman sublevel. Here we plot the detected population in the previously emptied hyperfine level versus the difference in frequency between the Raman lasers, centered on the 9.192 GHz ground state frequency. Atoms are pumped into the (a) $m=-3$ or (b) $m=+3$ Zeeman sublevel by driving a $\Delta m=-1$ transition using the Zeeman laser for (a) or a $\Delta m=+1$ transition for (b). This reversal is facilitated by inserting or removing a half-wave plate to change the handedness of the Zeeman beam from right to left circularly polarized. The inset plot has been vertically stretched to better illustrate the less extreme Zeeman sublevels.}
        \label{fig:RamanSpectra}
\end{centering}
\end{figure}

\section{Determination of $\langle m \rangle$}\label{sec:maverage}
As introduced in the previous section, the ground state cesium atoms are optically prepared prior to entering the interaction region.  We find that typically $\approx$~99\% of the ground state population is in the selected initial state, $(F,m) = (3,\pm3)$.  Since preparation into a single $m$ level is not perfect, careful determination of the average value of the $m$, denoted $\langle m \rangle$, is critical. (See Eq.~\eqref{eq:Reqslphaoverbetam} or \eqref{eq:ratio_corrected}.)  
We discuss the measurement of $\langle m \rangle$ and evaluate its uncertainty here.

We use Raman spectra like those shown in Fig.~\ref{fig:RamanSpectra} to determine the fractional population, denoted $f_m$, in each $m$ level.  We collect these spectra using a pair of 852~nm beams tuned to the Raman transition that couples the two hyperfine lines of the ground state.  
For the purposes of this discussion, we illustrate the process for the case in which we prepare the ground state atoms in the state $(F,m) = (3, 3)$, but the technique is applicable to other initial states as well.  The two lasers driving the Raman interaction are tuned to 852 nm, detuned from the $6s \rightarrow 6p_{3/2}$ transition by $\Delta \sim 2 \pi \times 1$ GHz.  The two lasers are phase-locked to one another, and frequency stabilized to a saturated absorption resonance in a cesium vapor cell. The beams are combined in a  polarization maintaining fiber patchcord and are collimated, linearly-polarized, and then circularly polarized. This polarization state drives $\Delta m = 0$ transitions, $(F, m) = (3, m) \rightarrow (4, m)$. The Raman and 540 nm beams are aligned such that they are spatially overlapped on the atomic beam, but with the Raman lasers blocked during ratio measurements and with the 540 nm and 1079 nm beams blocked during pumping efficiency measurements.  This allows direct detection of the population at the location of the intended interaction.  We drive the individual transitions by scanning the frequency difference between the Raman lasers and observe population that has been driven to the $F=4$ hyperfine level. We average fifteen scans across the transition. We show sample Raman spectra collected during the measurements in Fig.~\ref{fig:RamanSpectra}. 

Due to the requirement for a precise determination of $\langle m \rangle$, our measurements of $\alpha/\beta$ are restricted to the $F=3 \rightarrow F^{\prime}=3$ hyperfine line.  When pumping atoms to the $(F,m) = (4,\pm 4)$ ground state, a  $\Delta m = \pm 2$  Raman transition is necessary to determine $\langle m \rangle$.  The transition strength of the $\Delta m = \pm 2$ transitions are around 50 - 300 times weaker than that of the $\Delta m = 0$ transition. To determine $\langle m \rangle$ with better that 0.1$\%$ uncertainty, the polarization extinction ratio of the two Raman beams has to be beyond feasible to avoid driving $\Delta m = 0$ transitions.  Since the primary contribution of uncertainty in this measurement of $\alpha/\beta$ comes from the determination of $\langle m \rangle$ and $\Delta m = 0$ transitions would introduce a large systematic error, we chose to limit our measurements of the ratio $\alpha/\beta$ to the $6s, F=3 \rightarrow 7s, F^{\prime}=3$ transition. 
In light of recent work by Xiao \textit{et al.}~\cite{xiaoderevianko2023}, who evaluated corrections to the scalar and vector polarizabilities, as well as the magnitude of a tensor term, due to hyperfine coupling, and determined that these corrections are not observable at the current level of measurement sensitivities, the value of $\alpha/\beta$ is expected to be independent of which hyperfine line is used in the measurements.


The fractional population $f_m$ is proportional to $A_m/S_m$, where $A_m$ is the peak area and $S_m$ is the calculated Raman line strength.
We have determined that the weak $\Delta m = 0$ transitions are well below saturation levels, such that the peak area grows linearly with the square of the Rabi frequency for the transition, around $1-1.5\%$ of the atoms are excited.  
We observe reasonable agreement, $<7\%$ disagreement,  between the calculated line strengths, $S_m$, and the observed non-Zeeman pumped spectra. Here the atoms are only prepared into a single hyperfine level and are not deliberately pushed into the extreme Zeeman  levels.  Disagreement between calculated line strengths and the observed peak areas either originate due to the calculation of line strengths, from an uneven distribution of atoms in the Zeeman sublevels originating from the oven, and/or a weak Zeeman pumping effect due to the hyperfine pumping.  It is not critical to know the individual line strengths to $0.1\%$ if the pumping quality is sufficiently high such that the contribution of adjacent Zeeman levels is small.  We quantify $\langle m \rangle$ as

\begin{equation}
    \langle m \rangle =\sum^3_{m=-3} m f_m = \sum^3_{m=-3} m\frac{A_m}{S_m}
\end{equation}
where $A_m/S_m$ are normalized to achieve a total fraction  of one.  We measure $\langle m\rangle = 2.981(2)$, where $\langle m\rangle = 3.000$ would indicate perfect preparation in the $m=3$ state.  The deviation in $\langle m \rangle$ between preparation into the $m=+3$ or $m=-3$ is below 0.09\%.

\section{Systematic Contributions}\label{sec:SysContr}

It is critical to identify and reduce systematic effects in precision measurements of weak transitions. The largest systematic effect is that of the alignment of the polarization along the static electric field,  discussed in Sec.~\ref{sec:TPCC}.  To mitigate this effect, we measure and compare the modulation amplitudes with the electric field rotated $\pm 45^\circ$ from the vertical. If the polarization alignment is slightly rotated toward the $+45^\circ$ direction, for example, then the signal amplitude will be slightly larger when $\mathbf{E}$ is rotated to $+45^\circ$ than $-45^\circ$. We rotate the polarization $\boldsymbol{\varepsilon}^{2 \omega}$ of the green beam to equalize the signal modulation with $\mathbf{E}$ rotated to $+45^\circ$ or $-45^\circ$ to reduce $\varepsilon'_{x}/\varepsilon_{y}$ to less than $10^{-3}$ , where $\varepsilon'_{x}$ represents the real portion of the laser field perpendicular to the static electric field and $\varepsilon_{y}$ is the laser field parallel to the static electric field. This reduces the systematic uncertainty to below 90 ppm.  See Eq.~(\ref{eq:ratio_corrected}). An unwanted imaginary portion of the laser polarization,  $ \varepsilon_x''$, produces a systematic error that is first order in $\varepsilon''_{x}/\varepsilon_{y}$, but changes sign under an $m$ reversal. Using a crossed polarizer, we measure an extinction ratio of $2.5 \times 10^{-7}$, indicating that $\varepsilon_x^{\prime \prime} < 0.5 \times 10^{-3} \varepsilon_y $.   The effect of this circular polarization component $\varepsilon''_{x}$ causes a $\pm$0.8\% deviation in the ratio measurements under $\langle m \rangle$ reversal. This deviation is consistent with the measured circular polarization of the beam. We reduce this systematic effect by changing $m$ and averaging the ratios.

Other high order optical transitions must be minimized through careful alignment of static magnetic and electric fields along with polarization.  The ratio of moments for the magnetic dipole (M1) transition to the Stark vector transition is $M1/\beta\approx 29.5 $ V/cm \cite{BennettW99,antypasm12013}. 
We apply a static electric field of 430 V/cm and align the static electric and magnetic field such that the magnetic dipole transition drives a $\Delta m=\pm1$ transition while the vector Stark and two-photon transitions each drive a $\Delta m =0$ transition.
Perfect alignment of ~$\mathbf{B}$ along $\mathbf{k}$ would eliminate all two-photon/M1 interference and would not contribute to the systematic uncertainty.
We measure the transverse magnetic field components $B_x$ and $B_y$, by observing the Raman spectra, to be less than 10 mG, compared to $B_z \approx$ 8.8 G. This factor, along with the applied electric field and the ratio $M1/\beta\approx 29.5 $ V/cm, reduces the M1 contribution to below 80 ppm of the vector Stark amplitude.  The electric quadrupole transition is $\approx 230$ times smaller than the M1 transition \cite{BouchiatG88} and could potentially produce a $0.03\:\%$ contribution. In addition, this contribution is in phase with $\beta$ and does not reverse under an $\langle m \rangle$ reversal like $\beta$ does.  It, like the circular polarization error, averages out to zero. 

To search for and eliminate systematic contributions due to the applied electric field, we perform reversals of the applied static electric field as well as a rotation of the laser polarization by $90 ^\circ$. We see no variation with reversal of the electric field by $180 ^\circ$, and report the average of these results.  We also make measurements of the signal amplitude ratio upon reversal of the electric field, $0^\circ$ $\rightarrow$ $180^\circ$ from vertical. A ratio of one is expected in the absence of any stray fields. We see no significant deviation among the electric field reversals. The measurements with the laser polarization rotated by $90 ^\circ$ tests for any small ellipticity of the electrode ring pattern and resulted in a 0.15$\%$ deviation, as discussed in the next section.

\section{Results}\label{sec:Results}

The relative uncertainty in the measurement of the ratio $\mathcal{R}$ after 160 scans (1 hour) is typically $0.15-0.2\%$  and the reduced chi squared, $\chi^2_{red}$ ranges from 1 to 1.5.  For any data set for which $\chi^2_{red}>1$, we multiply the uncertainty by the square root of $\chi^2_{red}$. The primary contribution to this uncertainty is due to shot noise in the measurement and relative phase fluctuations between the 540 nm and 1079 nm beams. 

\begin{figure}
 \begin{centering}
 	  \includegraphics[width=0.48\textwidth]{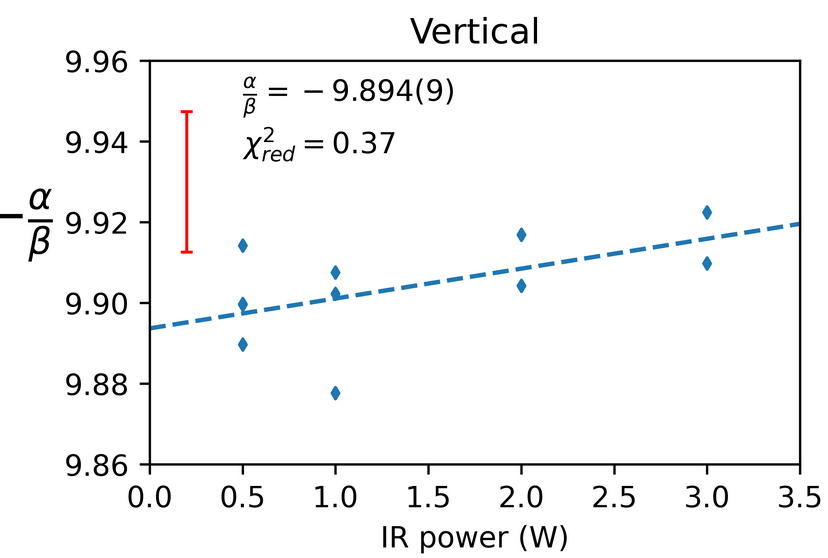}  \\ \vspace{-0.7in} \hspace{2.6in}{(a)  \hspace{0.9in}  } \hspace{0.1in} \\ \vspace{0.5in}
    \includegraphics[width=0.48\textwidth]{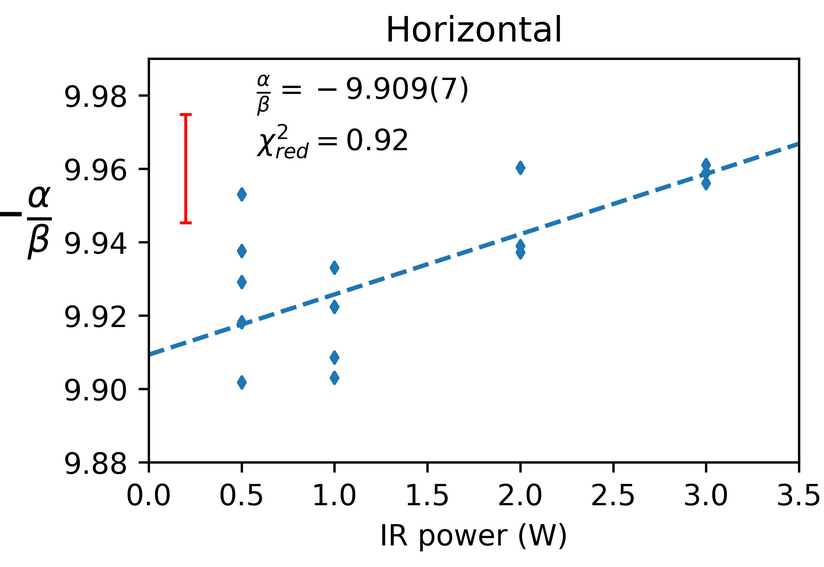}  \\ \vspace{-0.7in} \hspace{2.6in}{(b)  \hspace{0.9in}  }\\
 	  \vspace{4\baselineskip}
  \caption{Ratio measurements  plotted vs.~the IR laser power exciting the two-photon transition. The ratios are fitted to a straight line to determine the zero intensity ratio. The red error bars in each plot show the average uncertainty of the measured ratios (blue diamonds). Plot (a) shows the fitted ratio when the polarization is vertical and (b) shows the fitted ratio when the polarization is horizontal.}
        \label{fig:power_dependance}
\end{centering}
\end{figure}

To search for any possible ac Stark shifts that affect the ratio of $\alpha/\beta$, we collect scans at several different powers of the 1079 nm and 540 nm, and extrapolate to zero intensity.  We observe a slight dependence in the ratio of $\alpha/\beta$ on the 1097 nm beam intensity (See Fig. \ref{fig:power_dependance}.) and no deviation from the 540 nm beam intensity. We also carry out measurements with the polarization of the laser beams rotated from vertical to horizontal and see a minimal deviation,  $\alpha/\beta=-9.894\:(9)$ for vertical polarization and  $\alpha/\beta=-9.909\:(7)$ for horizontal. This difference could originate from a slightly elliptical shape of the electrode pattern, as discussed in Sec.~\ref{sec:SysContr}, or from statistical variations in the measurement of $\mathcal{R}$.  Due to the similar uncertainties and the reasons above, we report the unweighted average these ratios, $\alpha/\beta=-9.902\:(6)_{stat}(7)_{sys}$, as our final result, where the subscript `stat' indicates the uncertainty due to statistical fluctuations and the subscript `sys' indicates the systematic uncertainty, primarily due to the $\langle m \rangle$ determination. 


The primary sources of uncertainty  for these measurements result from statistical uncertainties of the least-squares fits to the sinusoidally-varying data, and from the determination of the average value of the magnetic sub-level $\langle m \rangle$.  
We tabulate these uncertainties  and their magnitudes, as well as other less significant uncertainties, in Table~\ref{table:sourcesoferror}.

\begin{figure}
\begin{center}
     \includegraphics[width=0.50\textwidth]{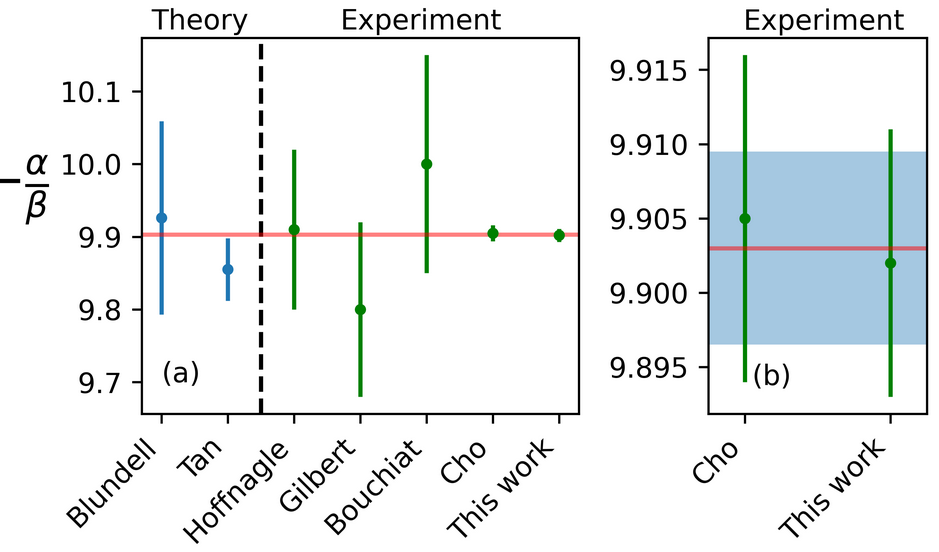}
 \caption{ Previous theoretical (left, blue) and experimental (right, green)  results of the ratio $\alpha/\beta$. Theoretical results are a sum-over-states calculations of $\alpha$ and $\beta$.  Plot (a) includes all previous results and plot (b) shows the two highest precision measurements on a finer scale. The pink horizontal line indicates the weighted average of the present result and that of Cho \textit{et al.}~\cite{ChoWBRW1997}, which we suggest as the recommended value, and the blue shaded region indicates the recommended uncertainty.}
 	  \label{fig:previousresults}
\end{center}
\end{figure}

We show past theoretical and experimental determinations of $\alpha/\beta$ in Fig.~\ref{fig:previousresults}.  The theoretical calculations use a sum-over-states approach to determine $\alpha$ and $\beta$ independently, and divide the results. Due to a large cancellation between terms of opposite sign in the calculation of $\beta$, the relative uncertainty in $\beta$ is much larger than in $\alpha$. For this reason, the theoretical calculations have difficulty attaining the same precision as experimental determinations, and  a measurement of $\alpha/\beta$ is critical for the determination of $\beta$. Fig.~\ref{fig:previousresults}(b) shows the present result and the previously accepted value of $\alpha/\beta$ by Cho \textit{et al.}~\cite{ChoWBRW1997}. Our measurement technique differs in several regards from that of Ref.~\cite{ChoWBRW1997}, including smaller influence of a.c.~Stark shifts, our use of two-color coherent control, and our use of much lower optical intensities and strictly linear field polarization. The two measured values are in excellent agreement. The pink line in (b) shows 
the weighted average of these two results, which we suggest as the recommended value, $\alpha/\beta=-9.903(6)$, and the shaded blue region indicates this recommended value's uncertainty. Using this recommended value of $\alpha/\beta$ and the sum-over-states calculation of $\alpha$ \cite{quirk2024starkshift}, we arrive at a new value for $\beta_\alpha=27.048\:(26)\: a_0^3$.   

\begin{table}

    \def\arraystretch{1.6}
    \centering
    \begin{tabular}{|@{\hskip 0.2in}l@{\hskip 0.5in}|@{\hskip 0.2in}c@{\hskip 0.3in}|}
    \hline
         source of uncertainty& $\sigma$ (ppm)\\
         \hline \hline \hline
         $\langle m \rangle$&670\\\hline
         fit&610\\\hline
         $\varepsilon_x'/\varepsilon_y$ (polarization alignment)& 90\\\hline
         $\varepsilon_x''/\varepsilon_y$(circular polarization)& 44\\\hline
         M1& 80\\\hline
         electric field& 50\\
 \hline \hline
 Total&920\\\hline
    \end{tabular}
    \caption{Sources and magnitudes of uncertainty for the determination of the ratio $\alpha / \beta$.  The primary sources of uncertainty originate from the fit and the determination of $\langle m \rangle$. We add the errors in quadrature to obtain the total uncertainty. }
    \label{table:sourcesoferror}
\end{table}

\section{Conclusion}\label{sec:Conclusion}
In this paper, we have described our new measurement of the ratio of the scalar to vector transition polarizability, $\alpha/\beta$. This precision measurement reaffirms the previously accepted value \cite{ChoWBRW1997} and  indicates a need to search along other avenues for the cause of the discrepancy between the two techniques that determine $\beta$. This discrepancy, $\beta_\alpha - \beta_{M1}=0.091\:(57)\: a_0^3$, must be resolved since $\beta$ is the moment to which the weak-force-induced electric dipole moment is scaled. 

\section{Acknowledgement}\label{sec:Acknowledgement}

We are grateful to A. Derevianko for helpful discussions, particularly on possible higher order polarizabilities. We would like to thank Aidan Jacobsen for his help in building the $\langle m \rangle$ reversal apparatus. This material is based upon work supported by the National Science Foundation under Grant Number PHY-1912519. 

\bibliography{biblio}

\begin{thebibliography}{37}%
\makeatletter
\providecommand \@ifxundefined [1]{%
 \@ifx{#1\undefined}
}%
\providecommand \@ifnum [1]{%
 \ifnum #1\expandafter \@firstoftwo
 \else \expandafter \@secondoftwo
 \fi
}%
\providecommand \@ifx [1]{%
 \ifx #1\expandafter \@firstoftwo
 \else \expandafter \@secondoftwo
 \fi
}%
\providecommand \natexlab [1]{#1}%
\providecommand \enquote  [1]{``#1''}%
\providecommand \bibnamefont  [1]{#1}%
\providecommand \bibfnamefont [1]{#1}%
\providecommand \citenamefont [1]{#1}%
\providecommand \href@noop [0]{\@secondoftwo}%
\providecommand \href [0]{\begingroup \@sanitize@url \@href}%
\providecommand \@href[1]{\@@startlink{#1}\@@href}%
\providecommand \@@href[1]{\endgroup#1\@@endlink}%
\providecommand \@sanitize@url [0]{\catcode `\\12\catcode `\$12\catcode `\&12\catcode `\#12\catcode `\^12\catcode `\_12\catcode `\%12\relax}%
\providecommand \@@startlink[1]{}%
\providecommand \@@endlink[0]{}%
\providecommand \url  [0]{\begingroup\@sanitize@url \@url }%
\providecommand \@url [1]{\endgroup\@href {#1}{\urlprefix }}%
\providecommand \urlprefix  [0]{URL }%
\providecommand \Eprint [0]{\href }%
\providecommand \doibase [0]{http://dx.doi.org/}%
\providecommand \selectlanguage [0]{\@gobble}%
\providecommand \bibinfo  [0]{\@secondoftwo}%
\providecommand \bibfield  [0]{\@secondoftwo}%
\providecommand \translation [1]{[#1]}%
\providecommand \BibitemOpen [0]{}%
\providecommand \bibitemStop [0]{}%
\providecommand \bibitemNoStop [0]{.\EOS\space}%
\providecommand \EOS [0]{\spacefactor3000\relax}%
\providecommand \BibitemShut  [1]{\csname bibitem#1\endcsname}%
\let\auto@bib@innerbib\@empty
\bibitem [{\citenamefont {Wood}\ \emph {et~al.}(1997)\citenamefont {Wood}, \citenamefont {Bennett}, \citenamefont {Cho}, \citenamefont {Masterson}, \citenamefont {Roberts}, \citenamefont {Tanner},\ and\ \citenamefont {Wieman}}]{WoodBCMRTW97}%
  \BibitemOpen
  \bibfield  {author} {\bibinfo {author} {\bibfnamefont {C.~S.}\ \bibnamefont {Wood}}, \bibinfo {author} {\bibfnamefont {S.~C.}\ \bibnamefont {Bennett}}, \bibinfo {author} {\bibfnamefont {D.}~\bibnamefont {Cho}}, \bibinfo {author} {\bibfnamefont {B.~P.}\ \bibnamefont {Masterson}}, \bibinfo {author} {\bibfnamefont {J.~L.}\ \bibnamefont {Roberts}}, \bibinfo {author} {\bibfnamefont {C.~E.}\ \bibnamefont {Tanner}}, \ and\ \bibinfo {author} {\bibfnamefont {C.~E.}\ \bibnamefont {Wieman}},\ }\href {\doibase 10.1126/science.275.5307.1759} {\bibfield  {journal} {\bibinfo  {journal} {Science}\ }\textbf {\bibinfo {volume} {275}},\ \bibinfo {pages} {1759} (\bibinfo {year} {1997})}\BibitemShut {NoStop}%
\bibitem [{\citenamefont {Dzuba}\ and\ \citenamefont {Flambaum}(2000)}]{DzubaF00}%
  \BibitemOpen
  \bibfield  {author} {\bibinfo {author} {\bibfnamefont {V.~A.}\ \bibnamefont {Dzuba}}\ and\ \bibinfo {author} {\bibfnamefont {V.~V.}\ \bibnamefont {Flambaum}},\ }\href {\doibase 10.1103/PhysRevA.62.052101} {\bibfield  {journal} {\bibinfo  {journal} {Phys. Rev. A}\ }\textbf {\bibinfo {volume} {62}},\ \bibinfo {pages} {052101} (\bibinfo {year} {2000})}\BibitemShut {NoStop}%
\bibitem [{\citenamefont {Bennett}\ and\ \citenamefont {Wieman}(1999)}]{BennettW99}%
  \BibitemOpen
  \bibfield  {author} {\bibinfo {author} {\bibfnamefont {S.~C.}\ \bibnamefont {Bennett}}\ and\ \bibinfo {author} {\bibfnamefont {C.~E.}\ \bibnamefont {Wieman}},\ }\href {\doibase 10.1103/PhysRevLett.82.2484} {\bibfield  {journal} {\bibinfo  {journal} {Physical Review Letters}\ }\textbf {\bibinfo {volume} {82}},\ \bibinfo {pages} {2484} (\bibinfo {year} {1999})}\BibitemShut {NoStop}%
\bibitem [{\citenamefont {Blundell}\ \emph {et~al.}(1992)\citenamefont {Blundell}, \citenamefont {Sapirstein},\ and\ \citenamefont {Johnson}}]{BlundellSJ92}%
  \BibitemOpen
  \bibfield  {author} {\bibinfo {author} {\bibfnamefont {S.~A.}\ \bibnamefont {Blundell}}, \bibinfo {author} {\bibfnamefont {J.}~\bibnamefont {Sapirstein}}, \ and\ \bibinfo {author} {\bibfnamefont {W.~R.}\ \bibnamefont {Johnson}},\ }\href {\doibase 10.1103/PhysRevD.45.1602} {\bibfield  {journal} {\bibinfo  {journal} {Phys. Rev. D}\ }\textbf {\bibinfo {volume} {45}},\ \bibinfo {pages} {1602} (\bibinfo {year} {1992})}\BibitemShut {NoStop}%
\bibitem [{\citenamefont {Safronova}\ \emph {et~al.}(1999)\citenamefont {Safronova}, \citenamefont {Johnson},\ and\ \citenamefont {Derevianko}}]{SafronovaJD99}%
  \BibitemOpen
  \bibfield  {author} {\bibinfo {author} {\bibfnamefont {M.~S.}\ \bibnamefont {Safronova}}, \bibinfo {author} {\bibfnamefont {W.~R.}\ \bibnamefont {Johnson}}, \ and\ \bibinfo {author} {\bibfnamefont {A.}~\bibnamefont {Derevianko}},\ }\href {\doibase 10.1103/PhysRevA.60.4476} {\bibfield  {journal} {\bibinfo  {journal} {Phys. Rev. A}\ }\textbf {\bibinfo {volume} {60}},\ \bibinfo {pages} {4476} (\bibinfo {year} {1999})}\BibitemShut {NoStop}%
\bibitem [{\citenamefont {Vasilyev}\ \emph {et~al.}(2002)\citenamefont {Vasilyev}, \citenamefont {Savukov}, \citenamefont {Safronova},\ and\ \citenamefont {Berry}}]{VasilyevSSB02}%
  \BibitemOpen
  \bibfield  {author} {\bibinfo {author} {\bibfnamefont {A.~A.}\ \bibnamefont {Vasilyev}}, \bibinfo {author} {\bibfnamefont {I.~M.}\ \bibnamefont {Savukov}}, \bibinfo {author} {\bibfnamefont {M.~S.}\ \bibnamefont {Safronova}}, \ and\ \bibinfo {author} {\bibfnamefont {H.~G.}\ \bibnamefont {Berry}},\ }\href {\doibase 10.1103/PhysRevA.66.020101} {\bibfield  {journal} {\bibinfo  {journal} {Phys. Rev. A}\ }\textbf {\bibinfo {volume} {66}},\ \bibinfo {pages} {020101(R)} (\bibinfo {year} {2002})}\BibitemShut {NoStop}%
\bibitem [{\citenamefont {Dzuba}\ \emph {et~al.}(2002)\citenamefont {Dzuba}, \citenamefont {Flambaum},\ and\ \citenamefont {Ginges}}]{DzubaFG02}%
  \BibitemOpen
  \bibfield  {author} {\bibinfo {author} {\bibfnamefont {V.~A.}\ \bibnamefont {Dzuba}}, \bibinfo {author} {\bibfnamefont {V.~V.}\ \bibnamefont {Flambaum}}, \ and\ \bibinfo {author} {\bibfnamefont {J.~S.~M.}\ \bibnamefont {Ginges}},\ }\href {\doibase 10.1103/PhysRevD.66.076013} {\bibfield  {journal} {\bibinfo  {journal} {Phys. Rev. D}\ }\textbf {\bibinfo {volume} {66}},\ \bibinfo {pages} {076013} (\bibinfo {year} {2002})}\BibitemShut {NoStop}%
\bibitem [{\citenamefont {Tran~Tan}\ \emph {et~al.}(2023)\citenamefont {Tran~Tan}, \citenamefont {Xiao},\ and\ \citenamefont {Derevianko}}]{TanXD2023}%
  \BibitemOpen
  \bibfield  {author} {\bibinfo {author} {\bibfnamefont {H.~B.}\ \bibnamefont {Tran~Tan}}, \bibinfo {author} {\bibfnamefont {D.}~\bibnamefont {Xiao}}, \ and\ \bibinfo {author} {\bibfnamefont {A.}~\bibnamefont {Derevianko}},\ }\href {\doibase 10.1103/PhysRevA.108.022808} {\bibfield  {journal} {\bibinfo  {journal} {Phys. Rev. A}\ }\textbf {\bibinfo {volume} {108}},\ \bibinfo {pages} {022808} (\bibinfo {year} {2023})}\BibitemShut {NoStop}%
\bibitem [{\citenamefont {Cho}\ \emph {et~al.}(1997)\citenamefont {Cho}, \citenamefont {Wood}, \citenamefont {Bennett}, \citenamefont {Roberts},\ and\ \citenamefont {Wieman}}]{ChoWBRW1997}%
  \BibitemOpen
  \bibfield  {author} {\bibinfo {author} {\bibfnamefont {D.}~\bibnamefont {Cho}}, \bibinfo {author} {\bibfnamefont {C.~S.}\ \bibnamefont {Wood}}, \bibinfo {author} {\bibfnamefont {S.~C.}\ \bibnamefont {Bennett}}, \bibinfo {author} {\bibfnamefont {J.~L.}\ \bibnamefont {Roberts}}, \ and\ \bibinfo {author} {\bibfnamefont {C.~E.}\ \bibnamefont {Wieman}},\ }\href {\doibase 10.1103/PhysRevA.55.1007} {\bibfield  {journal} {\bibinfo  {journal} {Phys. Rev. A}\ }\textbf {\bibinfo {volume} {55}},\ \bibinfo {pages} {1007} (\bibinfo {year} {1997})}\BibitemShut {NoStop}%
\bibitem [{\citenamefont {{Bouchiat, M.A.}}\ \emph {et~al.}(1984)\citenamefont {{Bouchiat, M.A.}}, \citenamefont {{Guena, J.}},\ and\ \citenamefont {{Pottier, L.}}}]{BouchiatGP84}%
  \BibitemOpen
  \bibfield  {author} {\bibinfo {author} {\bibnamefont {{Bouchiat, M.A.}}}, \bibinfo {author} {\bibnamefont {{Guena, J.}}}, \ and\ \bibinfo {author} {\bibnamefont {{Pottier, L.}}},\ }\href {\doibase 10.1051/jphyslet:019840045011052300} {\bibfield  {journal} {\bibinfo  {journal} {J. Physique Lett.}\ }\textbf {\bibinfo {volume} {45}},\ \bibinfo {pages} {523} (\bibinfo {year} {1984})}\BibitemShut {NoStop}%
\bibitem [{\citenamefont {Tanner}\ \emph {et~al.}(1992)\citenamefont {Tanner}, \citenamefont {Livingston}, \citenamefont {Rafac}, \citenamefont {Serpa}, \citenamefont {Kukla}, \citenamefont {Berry}, \citenamefont {Young},\ and\ \citenamefont {Kurtz}}]{TannerLRSKBYK92}%
  \BibitemOpen
  \bibfield  {author} {\bibinfo {author} {\bibfnamefont {C.~E.}\ \bibnamefont {Tanner}}, \bibinfo {author} {\bibfnamefont {A.~E.}\ \bibnamefont {Livingston}}, \bibinfo {author} {\bibfnamefont {R.~J.}\ \bibnamefont {Rafac}}, \bibinfo {author} {\bibfnamefont {F.~G.}\ \bibnamefont {Serpa}}, \bibinfo {author} {\bibfnamefont {K.~W.}\ \bibnamefont {Kukla}}, \bibinfo {author} {\bibfnamefont {H.~G.}\ \bibnamefont {Berry}}, \bibinfo {author} {\bibfnamefont {L.}~\bibnamefont {Young}}, \ and\ \bibinfo {author} {\bibfnamefont {C.~A.}\ \bibnamefont {Kurtz}},\ }\href {\doibase 10.1103/PhysRevLett.69.2765} {\bibfield  {journal} {\bibinfo  {journal} {Phys. Rev. Lett.}\ }\textbf {\bibinfo {volume} {69}},\ \bibinfo {pages} {2765} (\bibinfo {year} {1992})}\BibitemShut {NoStop}%
\bibitem [{\citenamefont {Young}\ \emph {et~al.}(1994)\citenamefont {Young}, \citenamefont {Hill}, \citenamefont {Sibener}, \citenamefont {Price}, \citenamefont {Tanner}, \citenamefont {Wieman},\ and\ \citenamefont {Leone}}]{YoungHSPTWL94}%
  \BibitemOpen
  \bibfield  {author} {\bibinfo {author} {\bibfnamefont {L.}~\bibnamefont {Young}}, \bibinfo {author} {\bibfnamefont {W.~T.}\ \bibnamefont {Hill}}, \bibinfo {author} {\bibfnamefont {S.~J.}\ \bibnamefont {Sibener}}, \bibinfo {author} {\bibfnamefont {S.~D.}\ \bibnamefont {Price}}, \bibinfo {author} {\bibfnamefont {C.~E.}\ \bibnamefont {Tanner}}, \bibinfo {author} {\bibfnamefont {C.~E.}\ \bibnamefont {Wieman}}, \ and\ \bibinfo {author} {\bibfnamefont {S.~R.}\ \bibnamefont {Leone}},\ }\href {\doibase 10.1103/PhysRevA.50.2174} {\bibfield  {journal} {\bibinfo  {journal} {Phys. Rev. A}\ }\textbf {\bibinfo {volume} {50}},\ \bibinfo {pages} {2174} (\bibinfo {year} {1994})}\BibitemShut {NoStop}%
\bibitem [{\citenamefont {Rafac}\ and\ \citenamefont {Tanner}(1998)}]{RafacT98}%
  \BibitemOpen
  \bibfield  {author} {\bibinfo {author} {\bibfnamefont {R.~J.}\ \bibnamefont {Rafac}}\ and\ \bibinfo {author} {\bibfnamefont {C.~E.}\ \bibnamefont {Tanner}},\ }\href {\doibase 10.1103/PhysRevA.58.1087} {\bibfield  {journal} {\bibinfo  {journal} {Phys. Rev. A}\ }\textbf {\bibinfo {volume} {58}},\ \bibinfo {pages} {1087} (\bibinfo {year} {1998})}\BibitemShut {NoStop}%
\bibitem [{\citenamefont {Rafac}\ \emph {et~al.}(1999)\citenamefont {Rafac}, \citenamefont {Tanner}, \citenamefont {Livingston},\ and\ \citenamefont {Berry}}]{RafacTLB99}%
  \BibitemOpen
  \bibfield  {author} {\bibinfo {author} {\bibfnamefont {R.~J.}\ \bibnamefont {Rafac}}, \bibinfo {author} {\bibfnamefont {C.~E.}\ \bibnamefont {Tanner}}, \bibinfo {author} {\bibfnamefont {A.~E.}\ \bibnamefont {Livingston}}, \ and\ \bibinfo {author} {\bibfnamefont {H.~G.}\ \bibnamefont {Berry}},\ }\href {\doibase 10.1103/PhysRevA.60.3648} {\bibfield  {journal} {\bibinfo  {journal} {Phys. Rev. A}\ }\textbf {\bibinfo {volume} {60}},\ \bibinfo {pages} {3648} (\bibinfo {year} {1999})}\BibitemShut {NoStop}%
\bibitem [{\citenamefont {Bennett}\ \emph {et~al.}(1999)\citenamefont {Bennett}, \citenamefont {Roberts},\ and\ \citenamefont {Wieman}}]{BennettRW99}%
  \BibitemOpen
  \bibfield  {author} {\bibinfo {author} {\bibfnamefont {S.~C.}\ \bibnamefont {Bennett}}, \bibinfo {author} {\bibfnamefont {J.~L.}\ \bibnamefont {Roberts}}, \ and\ \bibinfo {author} {\bibfnamefont {C.~E.}\ \bibnamefont {Wieman}},\ }\href {\doibase 10.1103/PhysRevA.59.R16} {\bibfield  {journal} {\bibinfo  {journal} {Phys. Rev. A}\ }\textbf {\bibinfo {volume} {59}},\ \bibinfo {pages} {R16} (\bibinfo {year} {1999})}\BibitemShut {NoStop}%
\bibitem [{\citenamefont {Derevianko}\ and\ \citenamefont {Porsev}(2002)}]{DereviankoP02a}%
  \BibitemOpen
  \bibfield  {author} {\bibinfo {author} {\bibfnamefont {A.}~\bibnamefont {Derevianko}}\ and\ \bibinfo {author} {\bibfnamefont {S.~G.}\ \bibnamefont {Porsev}},\ }\href {\doibase 10.1103/PhysRevA.65.053403} {\bibfield  {journal} {\bibinfo  {journal} {Phys. Rev. A}\ }\textbf {\bibinfo {volume} {65}},\ \bibinfo {pages} {053403} (\bibinfo {year} {2002})}\BibitemShut {NoStop}%
\bibitem [{\citenamefont {Amini}\ and\ \citenamefont {Gould}(2003)}]{AminiG03}%
  \BibitemOpen
  \bibfield  {author} {\bibinfo {author} {\bibfnamefont {J.~M.}\ \bibnamefont {Amini}}\ and\ \bibinfo {author} {\bibfnamefont {H.}~\bibnamefont {Gould}},\ }\href {\doibase 10.1103/PhysRevLett.91.153001} {\bibfield  {journal} {\bibinfo  {journal} {Phys. Rev. Lett.}\ }\textbf {\bibinfo {volume} {91}},\ \bibinfo {pages} {153001} (\bibinfo {year} {2003})}\BibitemShut {NoStop}%
\bibitem [{\citenamefont {Bouloufa}\ \emph {et~al.}(2007)\citenamefont {Bouloufa}, \citenamefont {Crubellier},\ and\ \citenamefont {Dulieu}}]{BouloufaCD07}%
  \BibitemOpen
  \bibfield  {author} {\bibinfo {author} {\bibfnamefont {N.}~\bibnamefont {Bouloufa}}, \bibinfo {author} {\bibfnamefont {A.}~\bibnamefont {Crubellier}}, \ and\ \bibinfo {author} {\bibfnamefont {O.}~\bibnamefont {Dulieu}},\ }\href {\doibase 10.1103/PhysRevA.75.052501} {\bibfield  {journal} {\bibinfo  {journal} {Phys. Rev. A}\ }\textbf {\bibinfo {volume} {75}},\ \bibinfo {pages} {052501} (\bibinfo {year} {2007})}\BibitemShut {NoStop}%
\bibitem [{\citenamefont {Sell}\ \emph {et~al.}(2011)\citenamefont {Sell}, \citenamefont {Patterson}, \citenamefont {Ehrenreich}, \citenamefont {Brooke}, \citenamefont {Scoville},\ and\ \citenamefont {Knize}}]{SellPEBSK11}%
  \BibitemOpen
  \bibfield  {author} {\bibinfo {author} {\bibfnamefont {J.~F.}\ \bibnamefont {Sell}}, \bibinfo {author} {\bibfnamefont {B.~M.}\ \bibnamefont {Patterson}}, \bibinfo {author} {\bibfnamefont {T.}~\bibnamefont {Ehrenreich}}, \bibinfo {author} {\bibfnamefont {G.}~\bibnamefont {Brooke}}, \bibinfo {author} {\bibfnamefont {J.}~\bibnamefont {Scoville}}, \ and\ \bibinfo {author} {\bibfnamefont {R.~J.}\ \bibnamefont {Knize}},\ }\href {\doibase 10.1103/PhysRevA.84.010501} {\bibfield  {journal} {\bibinfo  {journal} {Phys. Rev. A}\ }\textbf {\bibinfo {volume} {84}},\ \bibinfo {pages} {010501(R)} (\bibinfo {year} {2011})}\BibitemShut {NoStop}%
\bibitem [{\citenamefont {Zhang}\ \emph {et~al.}(2013)\citenamefont {Zhang}, \citenamefont {Ma}, \citenamefont {Wu}, \citenamefont {Wang}, \citenamefont {Xiao},\ and\ \citenamefont {Jia}}]{ZhangMWWXJ13}%
  \BibitemOpen
  \bibfield  {author} {\bibinfo {author} {\bibfnamefont {Y.}~\bibnamefont {Zhang}}, \bibinfo {author} {\bibfnamefont {J.}~\bibnamefont {Ma}}, \bibinfo {author} {\bibfnamefont {J.}~\bibnamefont {Wu}}, \bibinfo {author} {\bibfnamefont {L.}~\bibnamefont {Wang}}, \bibinfo {author} {\bibfnamefont {L.}~\bibnamefont {Xiao}}, \ and\ \bibinfo {author} {\bibfnamefont {S.}~\bibnamefont {Jia}},\ }\href {\doibase 10.1103/PhysRevA.87.030503} {\bibfield  {journal} {\bibinfo  {journal} {Phys. Rev. A}\ }\textbf {\bibinfo {volume} {87}},\ \bibinfo {pages} {030503(R)} (\bibinfo {year} {2013})}\BibitemShut {NoStop}%
\bibitem [{\citenamefont {Antypas}\ and\ \citenamefont {Elliott}(2013{\natexlab{a}})}]{antypas7p2013}%
  \BibitemOpen
  \bibfield  {author} {\bibinfo {author} {\bibfnamefont {D.}~\bibnamefont {Antypas}}\ and\ \bibinfo {author} {\bibfnamefont {D.~S.}\ \bibnamefont {Elliott}},\ }\href {\doibase 10.1103/PhysRevA.88.052516} {\bibfield  {journal} {\bibinfo  {journal} {Phys. Rev. A}\ }\textbf {\bibinfo {volume} {88}},\ \bibinfo {pages} {052516} (\bibinfo {year} {2013}{\natexlab{a}})}\BibitemShut {NoStop}%
\bibitem [{\citenamefont {Borv\'{a}k}(2014)}]{Borvak14}%
  \BibitemOpen
  \bibfield  {author} {\bibinfo {author} {\bibfnamefont {L.}~\bibnamefont {Borv\'{a}k}},\ }\emph {\bibinfo {title} {Direct laser absorption spectroscopy measurements of transition strengths in cesium}},\ \href@noop {} {Ph.D. thesis},\ \bibinfo  {school} {University of Notre Dame} (\bibinfo {year} {2014})\BibitemShut {NoStop}%
\bibitem [{\citenamefont {Patterson}\ \emph {et~al.}(2015)\citenamefont {Patterson}, \citenamefont {Sell}, \citenamefont {Ehrenreich}, \citenamefont {Gearba}, \citenamefont {Brooke}, \citenamefont {Scoville},\ and\ \citenamefont {Knize}}]{PattersonSEGBSK15}%
  \BibitemOpen
  \bibfield  {author} {\bibinfo {author} {\bibfnamefont {B.~M.}\ \bibnamefont {Patterson}}, \bibinfo {author} {\bibfnamefont {J.~F.}\ \bibnamefont {Sell}}, \bibinfo {author} {\bibfnamefont {T.}~\bibnamefont {Ehrenreich}}, \bibinfo {author} {\bibfnamefont {M.~A.}\ \bibnamefont {Gearba}}, \bibinfo {author} {\bibfnamefont {G.~M.}\ \bibnamefont {Brooke}}, \bibinfo {author} {\bibfnamefont {J.}~\bibnamefont {Scoville}}, \ and\ \bibinfo {author} {\bibfnamefont {R.~J.}\ \bibnamefont {Knize}},\ }\href {\doibase 10.1103/PhysRevA.91.012506} {\bibfield  {journal} {\bibinfo  {journal} {Phys. Rev. A}\ }\textbf {\bibinfo {volume} {91}},\ \bibinfo {pages} {012506} (\bibinfo {year} {2015})}\BibitemShut {NoStop}%
\bibitem [{\citenamefont {Gregoire}\ \emph {et~al.}(2015)\citenamefont {Gregoire}, \citenamefont {Hromada}, \citenamefont {Holmgren}, \citenamefont {Trubko},\ and\ \citenamefont {Cronin}}]{GregoireHHTC15}%
  \BibitemOpen
  \bibfield  {author} {\bibinfo {author} {\bibfnamefont {M.~D.}\ \bibnamefont {Gregoire}}, \bibinfo {author} {\bibfnamefont {I.}~\bibnamefont {Hromada}}, \bibinfo {author} {\bibfnamefont {W.~F.}\ \bibnamefont {Holmgren}}, \bibinfo {author} {\bibfnamefont {R.}~\bibnamefont {Trubko}}, \ and\ \bibinfo {author} {\bibfnamefont {A.~D.}\ \bibnamefont {Cronin}},\ }\href {\doibase 10.1103/PhysRevA.92.052513} {\bibfield  {journal} {\bibinfo  {journal} {Phys. Rev. A}\ }\textbf {\bibinfo {volume} {92}},\ \bibinfo {pages} {052513} (\bibinfo {year} {2015})}\BibitemShut {NoStop}%
\bibitem [{\citenamefont {Toh}\ \emph {et~al.}(2018)\citenamefont {Toh}, \citenamefont {Jaramillo-Villegas}, \citenamefont {Glotzbach}, \citenamefont {Quirk}, \citenamefont {Stevenson}, \citenamefont {Choi}, \citenamefont {Weiner},\ and\ \citenamefont {Elliott}}]{TohJGQSCWE18}%
  \BibitemOpen
  \bibfield  {author} {\bibinfo {author} {\bibfnamefont {G.}~\bibnamefont {Toh}}, \bibinfo {author} {\bibfnamefont {J.~A.}\ \bibnamefont {Jaramillo-Villegas}}, \bibinfo {author} {\bibfnamefont {N.}~\bibnamefont {Glotzbach}}, \bibinfo {author} {\bibfnamefont {J.}~\bibnamefont {Quirk}}, \bibinfo {author} {\bibfnamefont {I.~C.}\ \bibnamefont {Stevenson}}, \bibinfo {author} {\bibfnamefont {J.}~\bibnamefont {Choi}}, \bibinfo {author} {\bibfnamefont {A.~M.}\ \bibnamefont {Weiner}}, \ and\ \bibinfo {author} {\bibfnamefont {D.~S.}\ \bibnamefont {Elliott}},\ }\href {\doibase 10.1103/PhysRevA.97.052507} {\bibfield  {journal} {\bibinfo  {journal} {Phys. Rev. A}\ }\textbf {\bibinfo {volume} {97}},\ \bibinfo {pages} {052507} (\bibinfo {year} {2018})}\BibitemShut {NoStop}%
\bibitem [{\citenamefont {Toh}\ \emph {et~al.}(2019)\citenamefont {Toh}, \citenamefont {Damitz}, \citenamefont {Glotzbach}, \citenamefont {Quirk}, \citenamefont {Stevenson}, \citenamefont {Choi}, \citenamefont {Safronova},\ and\ \citenamefont {Elliott}}]{TohDGQSCSE19}%
  \BibitemOpen
  \bibfield  {author} {\bibinfo {author} {\bibfnamefont {G.}~\bibnamefont {Toh}}, \bibinfo {author} {\bibfnamefont {A.}~\bibnamefont {Damitz}}, \bibinfo {author} {\bibfnamefont {N.}~\bibnamefont {Glotzbach}}, \bibinfo {author} {\bibfnamefont {J.}~\bibnamefont {Quirk}}, \bibinfo {author} {\bibfnamefont {I.~C.}\ \bibnamefont {Stevenson}}, \bibinfo {author} {\bibfnamefont {J.}~\bibnamefont {Choi}}, \bibinfo {author} {\bibfnamefont {M.~S.}\ \bibnamefont {Safronova}}, \ and\ \bibinfo {author} {\bibfnamefont {D.~S.}\ \bibnamefont {Elliott}},\ }\href {\doibase 10.1103/PhysRevA.99.032504} {\bibfield  {journal} {\bibinfo  {journal} {Phys. Rev. A}\ }\textbf {\bibinfo {volume} {99}},\ \bibinfo {pages} {032504} (\bibinfo {year} {2019})}\BibitemShut {NoStop}%
\bibitem [{\citenamefont {Damitz}\ \emph {et~al.}(2019)\citenamefont {Damitz}, \citenamefont {Toh}, \citenamefont {Putney}, \citenamefont {Tanner},\ and\ \citenamefont {Elliott}}]{DamitzTPTE18a}%
  \BibitemOpen
  \bibfield  {author} {\bibinfo {author} {\bibfnamefont {A.}~\bibnamefont {Damitz}}, \bibinfo {author} {\bibfnamefont {G.}~\bibnamefont {Toh}}, \bibinfo {author} {\bibfnamefont {E.}~\bibnamefont {Putney}}, \bibinfo {author} {\bibfnamefont {C.~E.}\ \bibnamefont {Tanner}}, \ and\ \bibinfo {author} {\bibfnamefont {D.~S.}\ \bibnamefont {Elliott}},\ }\href@noop {} {\bibfield  {journal} {\bibinfo  {journal} {Physical Review A}\ }\textbf {\bibinfo {volume} {99}},\ \bibinfo {pages} {062510} (\bibinfo {year} {2019})}\BibitemShut {NoStop}%
\bibitem [{\citenamefont {Quirk}\ \emph {et~al.}(2024)\citenamefont {Quirk}, \citenamefont {Jacobsen}, \citenamefont {Damitz}, \citenamefont {Tanner},\ and\ \citenamefont {Elliott}}]{quirk2024starkshift}%
  \BibitemOpen
  \bibfield  {author} {\bibinfo {author} {\bibfnamefont {J.~A.}\ \bibnamefont {Quirk}}, \bibinfo {author} {\bibfnamefont {A.}~\bibnamefont {Jacobsen}}, \bibinfo {author} {\bibfnamefont {A.}~\bibnamefont {Damitz}}, \bibinfo {author} {\bibfnamefont {C.~E.}\ \bibnamefont {Tanner}}, \ and\ \bibinfo {author} {\bibfnamefont {D.~S.}\ \bibnamefont {Elliott}},\ }\href@noop {} {\  (\bibinfo {year} {2024})},\ \Eprint {http://arxiv.org/abs/2311.09169} {arXiv:2311.09169 [physics.atom-ph]} \BibitemShut {NoStop}%
\bibitem [{\citenamefont {Tran~Tan}\ and\ \citenamefont {Derevianko}(2023)}]{TanD2023}%
  \BibitemOpen
  \bibfield  {author} {\bibinfo {author} {\bibfnamefont {H.~B.}\ \bibnamefont {Tran~Tan}}\ and\ \bibinfo {author} {\bibfnamefont {A.}~\bibnamefont {Derevianko}},\ }\href {\doibase 10.1103/PhysRevA.107.042809} {\bibfield  {journal} {\bibinfo  {journal} {Phys. Rev. A}\ }\textbf {\bibinfo {volume} {107}},\ \bibinfo {pages} {042809} (\bibinfo {year} {2023})}\BibitemShut {NoStop}%
\bibitem [{\citenamefont {Antypas}\ and\ \citenamefont {Elliott}(2013{\natexlab{b}})}]{antypasm12013}%
  \BibitemOpen
  \bibfield  {author} {\bibinfo {author} {\bibfnamefont {D.}~\bibnamefont {Antypas}}\ and\ \bibinfo {author} {\bibfnamefont {D.~S.}\ \bibnamefont {Elliott}},\ }\href {\doibase 10.1103/PhysRevA.87.042505} {\bibfield  {journal} {\bibinfo  {journal} {Phys. Rev. A}\ }\textbf {\bibinfo {volume} {87}},\ \bibinfo {pages} {042505} (\bibinfo {year} {2013}{\natexlab{b}})}\BibitemShut {NoStop}%
\bibitem [{\citenamefont {Antypas}\ and\ \citenamefont {Elliott}(2014)}]{antypase2014}%
  \BibitemOpen
  \bibfield  {author} {\bibinfo {author} {\bibfnamefont {D.}~\bibnamefont {Antypas}}\ and\ \bibinfo {author} {\bibfnamefont {D.}~\bibnamefont {Elliott}},\ }\href@noop {} {\bibfield  {journal} {\bibinfo  {journal} {Canadian Journal of Chemistry}\ }\textbf {\bibinfo {volume} {92}},\ \bibinfo {pages} {144} (\bibinfo {year} {2014})}\BibitemShut {NoStop}%
\bibitem [{\citenamefont {Bonin}\ and\ \citenamefont {McIlrath}(1984)}]{bonin1984two}%
  \BibitemOpen
  \bibfield  {author} {\bibinfo {author} {\bibfnamefont {K.~D.}\ \bibnamefont {Bonin}}\ and\ \bibinfo {author} {\bibfnamefont {T.~J.}\ \bibnamefont {McIlrath}},\ }\href@noop {} {\bibfield  {journal} {\bibinfo  {journal} {JOSA B}\ }\textbf {\bibinfo {volume} {1}},\ \bibinfo {pages} {52} (\bibinfo {year} {1984})}\BibitemShut {NoStop}%
\bibitem [{\citenamefont {Gilbert}\ and\ \citenamefont {Wieman}(1986)}]{GilbertW1986}%
  \BibitemOpen
  \bibfield  {author} {\bibinfo {author} {\bibfnamefont {S.}~\bibnamefont {Gilbert}}\ and\ \bibinfo {author} {\bibfnamefont {C.}~\bibnamefont {Wieman}},\ }\href@noop {} {\bibfield  {journal} {\bibinfo  {journal} {Physical Review A}\ }\textbf {\bibinfo {volume} {34}},\ \bibinfo {pages} {792} (\bibinfo {year} {1986})}\BibitemShut {NoStop}%
\bibitem [{\citenamefont {Wood}(1996)}]{Woodthesis1996}%
  \BibitemOpen
  \bibfield  {author} {\bibinfo {author} {\bibfnamefont {C.~S.}\ \bibnamefont {Wood}},\ }\emph {\bibinfo {title} {High precision atomic parity non-conservation measurement using a spin polarized Cesium beam and the uclear anapole moment of $^{133}$Cs}},\ \href@noop {} {Ph.D. thesis},\ \bibinfo  {school} {University of Colorado} (\bibinfo {year} {1996})\BibitemShut {NoStop}%
\bibitem [{\citenamefont {Antypas}(2013)}]{dionysiosthesis2013}%
  \BibitemOpen
  \bibfield  {author} {\bibinfo {author} {\bibfnamefont {D.}~\bibnamefont {Antypas}},\ }\emph {\bibinfo {title} {Measurement of a weak transition moment using coherent control}},\ \href@noop {} {Ph.D. thesis},\ \bibinfo  {school} {Purdue University} (\bibinfo {year} {2013})\BibitemShut {NoStop}%
\bibitem [{\citenamefont {Xiao}\ \emph {et~al.}(2023)\citenamefont {Xiao}, \citenamefont {Tan},\ and\ \citenamefont {Derevianko}}]{xiaoderevianko2023}%
  \BibitemOpen
  \bibfield  {author} {\bibinfo {author} {\bibfnamefont {D.}~\bibnamefont {Xiao}}, \bibinfo {author} {\bibfnamefont {H.~B.~T.}\ \bibnamefont {Tan}}, \ and\ \bibinfo {author} {\bibfnamefont {A.}~\bibnamefont {Derevianko}},\ }\href {\doibase 10.1103/PhysRevA.108.032805} {\bibfield  {journal} {\bibinfo  {journal} {Phys. Rev. A}\ }\textbf {\bibinfo {volume} {108}},\ \bibinfo {pages} {032805} (\bibinfo {year} {2023})}\BibitemShut {NoStop}%
\bibitem [{\citenamefont {Bouchiat}\ and\ \citenamefont {Gu{\'e}na}(1988)}]{BouchiatG88}%
  \BibitemOpen
  \bibfield  {author} {\bibinfo {author} {\bibfnamefont {M.-A.}\ \bibnamefont {Bouchiat}}\ and\ \bibinfo {author} {\bibfnamefont {J.}~\bibnamefont {Gu{\'e}na}},\ }\href@noop {} {\bibfield  {journal} {\bibinfo  {journal} {Journal de Physique}\ }\textbf {\bibinfo {volume} {49}},\ \bibinfo {pages} {2037} (\bibinfo {year} {1988})}\BibitemShut {NoStop}%
\end{thebibliography}%

\end{document}